\documentclass[12pt]{article}

\usepackage{newtxtext,newtxmath}
\usepackage{graphicx}

\usepackage[letterpaper,margin=1in]{geometry}

\renewenvironment{abstract}
	{\quotation}
	{\endquotation}

\date{}

\makeatletter
\renewcommand{\fnum@figure}{\textbf{Figure \thefigure}}
\renewcommand{\fnum@table}{\textbf{Table \thetable}}
\makeatother

\usepackage{scicite}

\usepackage{url}
\usepackage{graphicx}%
\usepackage{multirow}%
\usepackage{mathrsfs}%
\usepackage[title]{appendix}%
\usepackage{xcolor}%
\usepackage{textcomp}%
\usepackage{manyfoot}%
\usepackage{booktabs}%
\usepackage{algorithm}%
\usepackage{algorithmicx}%
\usepackage{algpseudocode}%
\usepackage{listings}%
\usepackage{gensymb}
\usepackage{color,soul}
\usepackage{float}

\def\scititle{
	Atomic short-range order: a new degree of freedom for band engineering of GeSn semiconductor alloys
}
\title{\bfseries \boldmath \scititle}

\author{
    Shang~Liu$^{1\dagger}$,
	Yunfan~Liang$^{2\dagger}$,
	Nirosh~M.~Eldose$^{3\dagger}$,
    Shunda~Chen$^{4}$,
    Xiaochen~Jin$^{4}$,\and
    Haochen~Zhao$^{5}$,
    Manoj~Shah$^{6}$,
    Jin-Hee~Bae$^{7}$,
    Omar~Concepcion$^{7}$,\and
    Fernando~M.~de Oliveira$^{3}$,
    Ilias~Bikmukhametov$^{8}$,
    Xiaoxin~Wang$^{1}$,
    Yuping~Zeng$^{5}$,\and
    Dan~Buca$^{7}$, 
    Mansour~Mortazavi$^{6}$,
    Damien~West$^{2}$, 
    Shengbai~Zhang$^{2}$,
    Tianshu~Li$^{4}$,\and
    Gregory~J.~Salamo$^{3}$,
    Shui-Qing~Yu$^{3,9}$,
    Jifeng~Liu$^{1\ast}$ \and
	\small$^{1}$Thayer School of Engineering, Dartmouth College, Hanover, NH 03755, USA.\and
    \small$^{2}$Department of Physics, Applied Physics and Astronomy, Rensselaer Polytechnic Institute, Troy, NY 12180, USA.\and
    \small$^{3}$Institute for Nanoscience and Engineering, University of Arkansas, Fayetteville, AR 72701, USA.\and
     \small$^{4}$Department of Civil and Environmental Engineering, George Washington University, Washington, DC 20052, USA.\and
    \small$^{5}$Department of Electrical and Computer Engineering, University of Delaware, Newark, DE 19716, USA.\and
    \small$^{6}$Department of Chemistry and Phyics, University of Arkansas, Pine Bluff, AR 71601, USA.\and
    \small$^{7}$Peter Gruenberg Institute 9 (PGI-9), Forschungszentrum Juelich, Juelich, 52428, Germany.\and
       \small$^{8}$Department of Metallurgical \& Materials Engineering, University of Alabama, Tuscaloosa, AL 35405, USA.\and
      \small$^{9}$Department of Electrical Engineering and Computer Science, University of Arkansas, Fayetteville, AR 72701, USA.\and
	\small$^\ast$Corresponding author. Email: Jifeng.Liu@dartmouth.edu\and
	\small$^\dagger$These authors contributed equally to this work.
}

\begin{document} 

% Insert the title and author list
\maketitle

% Abstract, in bold
% There are strict length limits, and not all formats have abstracts.
% Consult the journal instructions to authors for details.
% Do not cite any references in the abstract.
\begin{abstract} \bfseries \boldmath
% Start with one or two sentences of background

Chemical short-range order (SRO) in alloys denotes the statistical preference or avoidance between atomic species on neighboring lattice sites. Here, we highlight SRO as a powerful new mechanism for semiconductor alloy band engineering. Atom probe tomography reveals a significantly higher probability of Sn-Sn first nearest neighbors (1NNs) in thin-film GeSn alloys grown by molecular beam epitaxy (MBE) vs. chemical vapor deposition (CVD). Remarkably, although lower Sn concentration typically widens the bandgap, we find that the stronger presence of Sn-Sn 1NN pairs in MBE samples \textit{overrides} this trend, resulting in a narrower bandgap despite having 2 at.\% lower Sn content than CVD samples. First-principles modeling corroborates this effect, attributing these SRO variations to distinctive surface terminations and growth temperatures between MBE and CVD. These findings establish SRO as a new degree of freedom for semiconductor band engineering beyond composition, strain, and quantum confinement, unlocking novel device mechanisms for the post-Moore era. 

\end{abstract}

%% main text
\section{Introduction}
\label{sec:sample1}
Chemical short-range order (SRO)—the statistical preference or repulsion between atomic species on neighboring lattice sites—has recently emerged as a major focus in alloy research. On one hand, SRO has been found to affect mechanical properties of metallic medium/high entropy alloys (MEAs/HEAs) \cite{zhang2020short,chen2021direct,walsh2023extra, chen2021simultaneously,yin2021atomistic}.
On the other hand, SRO has been theoretically predicted to significantly impact the electronic band structure and thermal conductivity of semiconductor alloys such as GeSn and SiGeSn \cite{jin2022coexistence,jin2023role,cao2020short,chen2023impact,chen2024intricate}, attractive candidates for Si-compatible direct-bandgap infrared optoelectronics \cite{wirths2015lasing,homewood2015rise,kim2023short,zhou2022electrically,margetis2017si,talamas2021cmos,tran2019si,chen2023route}. A depletion of Sn–Sn 1$^{st}$ nearest neighbors (1NN) is predicted in GeSn alloys under thermodynamic equilibrium, leading to $\sim$0.1 eV, $\sim$0.16 eV and $\sim$0.33 eV larger direct bandgaps in Ge$_{0.875}$Sn$_{0.125}$, Ge$_{0.8125}$Sn$_{0.1875}$ and Ge$_{0.75}$Sn$_{0.25}$, respectively, compared to random alloys of the same compositions where Ge and Sn atoms randomly occupy the lattice sites \cite{cao2020short,chen2024intricate}. In recent years, various types of experimental analyses of SRO in GeSn and SiGeSn semiconductor alloys have been conducted. Extended X-ray Absorption Fine Structure (EXAFS) studies on GeSn \cite{gencarelli2015extended,lentz2023local, JAPPerspective, soo2014substitutional,EXAFS2025-2} have confirmed the tendency of Sn-Sn 1NN repulsion compared to random alloys. Furthermore, a very recent breakthrough in energy-filtered four-dimensional scanning transmission electron microscopy (EF-4D-STEM) has successfully identified the SRO motifs in SiGeSn alloys for the first time {\cite{vogl2024explore, vogl2024identification}}. 

We envision that harnessing SRO in semiconductor alloys will unlock new device mechanisms for the post-Moore era. However, \textit{experimentally} controlling SRO and verifying its impact on semiconductor alloy properties remain largely unexplored. Comparing SRO across samples grown under different conditions is indispensable for such studies, yet it poses great challenges. In the case of GeSn, EXAFS is able to provide the mean SRO parameter \cite{lentz2023local, EXAFS2025-1,EXAFS2025-2}, but incapable of probing the spatial distribution of SRO at nanoscale due to the X-ray beam size limitation. Such SRO spatial distribution also impacts the band structure on top of the mean value \cite{jin2023role,chen2024intricate}, therefore important for SRO comparison. State-of-the-art EF-4D-STEM offers the best spatial resolution for SRO mapping at nanoscale, yet it observed only weak diffuse diffraction signals from GeSn that are characteristic of SRO, in contrast to SiGeSn ternary alloys {\cite{vogl2024explore, vogl2024identification}}. In principle, the disorder-activated mode (DA) peak around 280 cm$^{-1}$ in Raman spectroscopy could be a facile probe of local ordering induced by Sn atoms based on first-principles theoretical modeling \mbox{\cite{corley2024polarization,corley2023local}}. However, the DA peak appears as a small shoulder partially overlapping the strong Ge-Ge peak, rendering it difficult to isolate the DA peak for SRO comparison. 

To address the challenges in SRO comparison discussed above, here we employ atom probe tomography (APT). This technique measures SRO parameters in real space \cite{li2023quantitative, liu20223d, he2024quantifying} and provides nano-scale mapping of SRO upon refined data analyses\cite{liu20223d}. Although atomic position perturbations limit the accuracy of APT in retrieving absolute SRO parameters, we have developed a physics-informed Poisson-KNN statistical method to reconstruct the K$^{th}$ nearest neighbor (KNN) shell from the raw APT data \cite{liu20223d}. This approach enables a reliable \textit{relative} comparison of SRO parameters across different samples. Using this method, we compare the SRO in GeSn grown by molecular beam epitaxy (MBE) vs. chemical vapor deposition (CVD)—the two dominant semiconductor alloy growth techniques\cite{zhang2020structural,wang2020spontaneously,oehme2012gesn,miao2021review,margetis2017study,dou2018crystalline}—and experimentally reveal its dramatic impact on the bandgap. A higher likelihood of Sn-Sn 1NNs is observed in GeSn grown by MBE at low temperatures (120-150  \degree C), consistent with a recent Density Functional Theory (DFT) modeling study on GeSn MBE growth \cite{Liang2025}. 

Remarkably, although lower Sn concentration usually widens the bandgap, we find that the stronger presence of Sn-Sn 1NN pairs in MBE samples \textit{overrides} this trend, resulting in a narrower bandgap despite having 2 at.\% lower Sn content than CVD samples. First-principles modeling attributes these SRO variations to distinctions in surface terminations and growth temperatures between MBE and CVD. Comparison between experimental and theoretical analyses indicates that increasing Sn-Sn 1NN SRO parameter by $\sim$0.7 would decrease the bandgap by at least 85 meV, equivalent to the impact of increasing Sn composition by $\sim$3 at.\%. This significant redshift represents $\sim$20\% of the material's bandgap ($\sim$500 meV). These findings establish SRO as a new degree of freedom for semiconductor band engineering beyond composition, strain, and quantum confinement. Because this new paradigm only alters the atomic neighborhood without changing macroscopic chemistry, crystal structure, or lattice constant, it unlocks novel SRO-based heterostructures and reconfigurable electronic/photonic materials that overcome conventional constraints for the post-Moore era. 

\section{Results and discussion}\label{sec2}

\subsection{APT data}\label{subsec1}

APT data of four GeSn samples have been analyzed. Samples \#1 and \#2 were grown by MBE, with 20 at.\% and 7 at.\% Sn, by two different tools at University of Delaware and University of Arkansas \cite{Eldose_2025}, respectively. The MBE growth temperature was 120-150  \degree C. Samples \#3 and \#4 were grown by CVD, with 14 at.\% and 7 at.\% Sn, respectively. They were grown by two different CVD reactors at University of Arkansas and Forschungszentrum Juelich, respectively. The corresponding growth temperature window was 250-350 \degree C, much higher than MBE due to the requirement of thermal decomposition of CVD precursors. Moreover, samples \#1 and \#3 are GeSn thin films, while samples \#2 and \#4 are GeSn/Ge multiple quantum wells (MQWs). The growth method and atomic composition are tabulated in Table \ref{tbl:Composition}. The APT measurement conditions are summarized in Table S1 of the Supporting Information.

\begin{table}[!htb]
\aboverulesep=0ex 
\belowrulesep=0ex
\caption{Growth method and composition}
  \label{tbl:Composition}
  \centering
\begin{tabular}{|c|c|c|c|c|}
\hline
Sample \# & Growth method                                   & Ge (at.\%) & Sn (at.\%) & Structure \\ \hline
1         & \multirow{2}{*}{MBE}  & 80       & 20       & Thin film \\ \cmidrule{1-1} \cmidrule{3-5} 
2         &                                                  & 93       & 7       & MQW \\ \hline
3         & \multirow{2}{*}{CVD} & 86       & 14       & Thin film \\ \cmidrule{1-1} \cmidrule{3-5} 
4         &                                                  & 93       & 7       & MQW \\ \hline
\end{tabular}
\end{table}

APT data of sample \#1, \#2 and \#4 are presented in Fig. \ref{fig:APT_Tips}. Fig. \ref{fig:APT_Tips}a, d, g are composition profiles of these three samples derived from their APT data; Fig. \ref{fig:APT_Tips}b, e, h are projections of reconstructed 3D atomic positions; Fig. \ref{fig:APT_Tips}c, f, i are spatial distribution maps of Sn-Sn 1NN SRO parameters which will be explained in the next section. APT data of sample \#3 has been published in Ref. \cite{liu20223d} and therefore not repeated here.

\subsection{Sn-Sn SRO parameter analyses and comparison }\label{subsec2}

The experimental SRO parameter is defined as:
\begin{equation}
  \alpha^{KNN}_{A-B} = \frac{P^{KNN}_{A-B}}{x_{B}}
    \begin{cases}
      >1, & B \;atoms \;are \;favored \;in \;the \;K^{th} \;shell \;of \;A \;atom  \\
      =1, & random \;alloy \\
      <1, & B \;atoms \;are \;not \;favored \;in \;the \;K^{th} \;shell \;of \;A \;atom 
    \end{cases},
  \label{eqn:SRO_Parameter_KNN}
\end{equation}
where $P^{KNN}_{A-B}$ denotes the statistical probability that a B atom is a KNN of an A atom, and $x_{B}$ represents the atomic fraction of B within the analyzed region (A,B=Ge,Sn). For a GeSn random alloy, $\alpha^{KNN}_{Sn-Sn}=1$. On the contrary, $\alpha^{KNN}_{Sn-Sn}>1$ (or $<1$) means a higher (or lower) likelihood for an Sn atom have other Sn atoms in its KNN shell compared to a random alloy of the same composition.

Note that our definition of SRO parameter in Equation (1) is exactly equal to 1 minus the Warren-Cowley SRO parameter commonly used in theoretical literature. Here, we use the experimental definition in Equation (1) because it intuitively correlates larger A-B KNN SRO parameters with higher likelihood of A-B KNN pairs.

\begin{figure}[H]\centering
    \center{\includegraphics[width=0.93\textwidth]
    {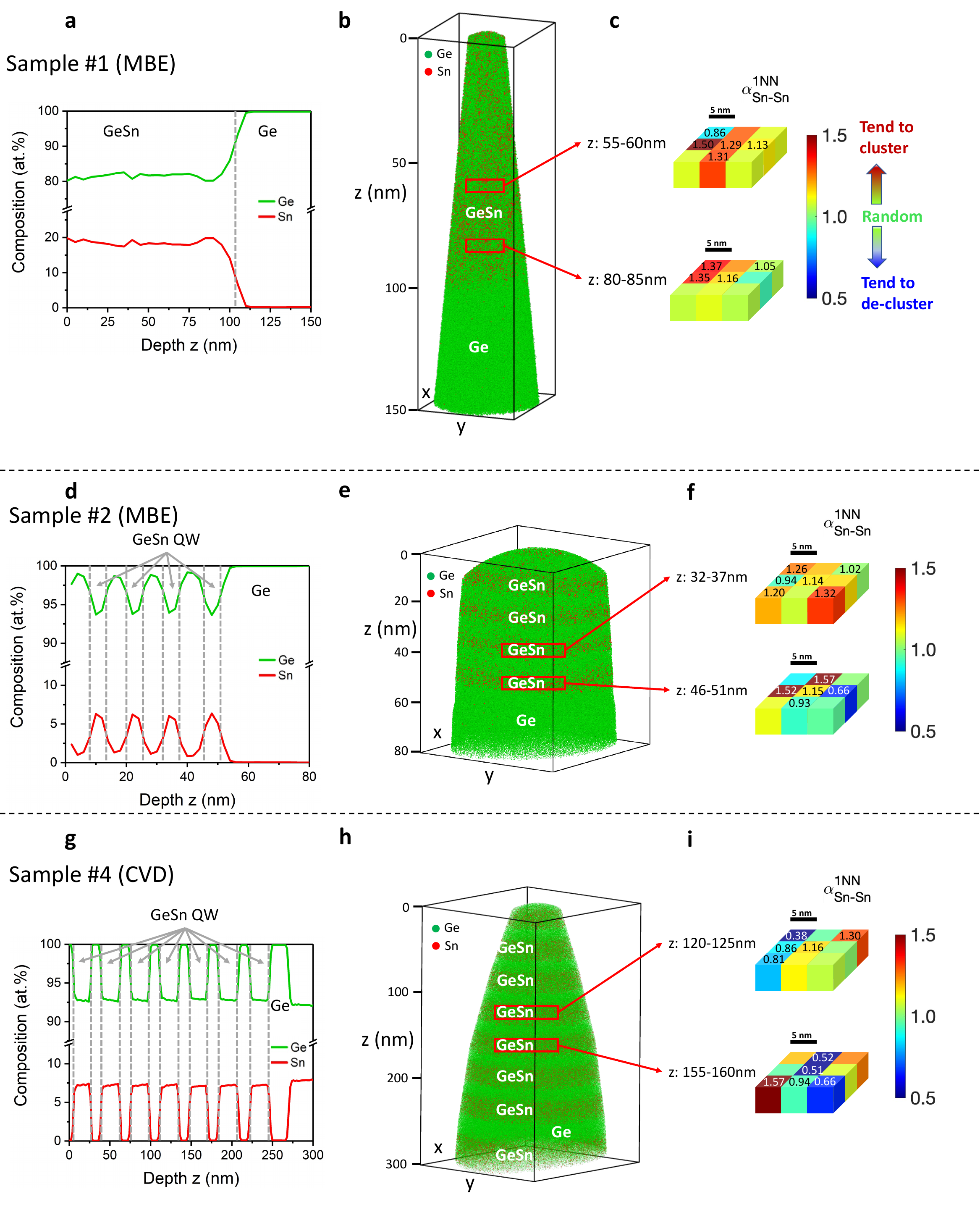}}
    \caption {APT data. a) Composition profile in sample \#1. b) Reconstructed APT data of sample \#1. c) SRO mapping examples in GeSn region in sample \#1. d) Composition profile in sample \#2. e) Reconstructed APT data of sample \#2. f) SRO mapping examples in GeSn region in sample \#2. g) Composition profile in sample \#4. h) Reconstructed APT data of sample \#4. i) SRO mapping examples in GeSn region in sample \#4.}
    \label{fig:APT_Tips}
\end{figure}

In order to address atomic position perturbations induced by field-evaporation in the APT data, the Poisson-KNN method developed in Ref.\cite{liu20223d} is used to reconstruct the KNN shells by a statistically weighted sum of nominal KNN in the APT data. More details are presented in the Experimental Section. We test the robustness of this Poisson-KNN method in distinguishing SRO from random alloys in the presence of atomic position perturbations, as detailed in Section 2 of the Supporting Information. This analysis shows that the Poisson-KNN method can successfully distinguish a random $Ge_{0.75}$$Sn_{0.25}$ alloy from a SRO $Ge_{0.75}$$Sn_{0.25}$ alloy with $\alpha^{1NN}_{Sn-Sn}$=0.22 for a standard deviation of 3 \text{\AA} in random atomic displacement, although  the difference in SRO parameters tends to be underestimated as the atomic position perturbation increases (see Fig. S2a in the Supporting Information). Therefore, the difference in SRO parameters between different samples derived by Poisson-KNN method from the corresponding APT data should be considered as the \textit{lower limit}. We further confirm that the standard deviation of reconstructed 1NN interatomic distance distribution in all four sets of APT data are almost identical ($\sim$1.0 \AA) with a variation smaller than 0.1 \AA \ among them (see Fig. S2b in the Supporting Information), indicating that the atomic perturbation is almost identical across all 4 samples. Therefore, the relative comparison between the MBE and CVD SRO parameters using Poisson-KNN method is well validated.

\begin{figure}[H]\centering
    \center{\includegraphics[width=\textwidth]
    {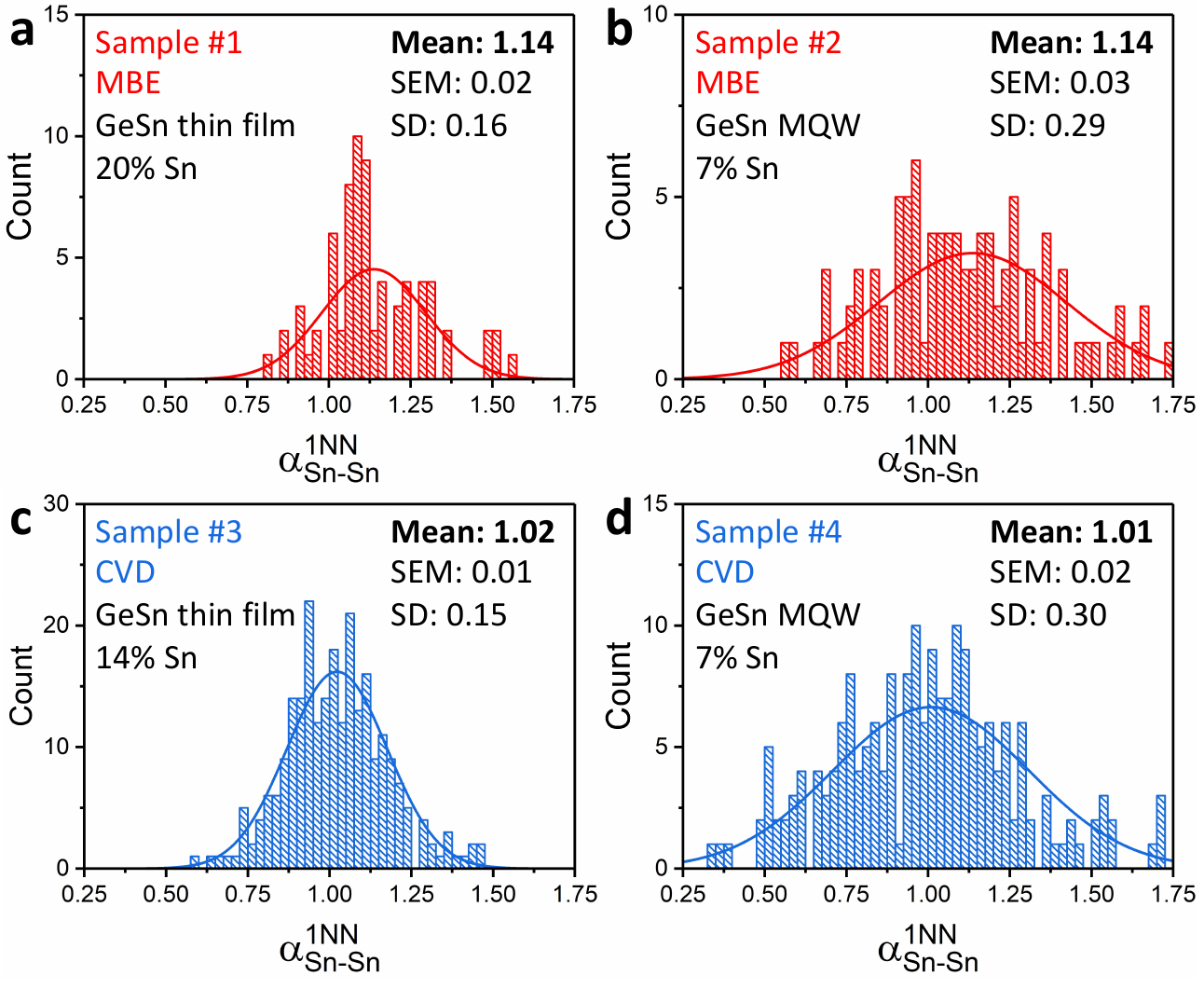}}
    \caption{Histogram of Sn-Sn 1NN SRO $\alpha^{1NN}_{Sn-Sn}$ in four samples. SEM: standard error of the mean. SD: standard deviation. Sample size = 72, 100, 250, and 200 (nanocubes) for these four samples. A Welch’s t-test reveals a highly statistically significant difference in the mean Sn-Sn 1NN SRO parameters between the MBE vs. CVD samples (p$<$0.0001)}
    \label{fig:SnSn_SRO_Histogram}
\end{figure}

Sn-Sn 1NN SRO distribution maps with a spatial resolution of 5 nm $\times$ 5 nm $\times$ 5 nm are produced and exemplified in Figs. \ref{fig:APT_Tips} c, f, and i. More cubes with $\alpha^{1NN}_{Sn-Sn}$ larger than 1.2 can be found in sample \#1 and \#2, while more cubes with $\alpha^{1NN}_{Sn-Sn}$ lower than 0.8 can be found in sample \#4. Therefore, qualitatively, MBE samples show more preference of Sn-Sn 1NN than CVD samples. Quantitatively, histograms of the Sn-Sn 1NN SRO $\alpha^{1NN}_{Sn-Sn}$ in these four samples are further compared in Fig. \ref{fig:SnSn_SRO_Histogram}. The red color is used for MBE samples, and blue color represents CVD samples. In these four samples, 72, 100, 250, and 200 nanocubes were analyzed, mainly limited by the thickness of each sample. The mean value, standard error of the mean (SEM), and standard deviation (SD) of $\alpha^{1NN}_{Sn-Sn}$ are labeled next to each histogram. Note that $\alpha^{1NN}_{Sn-Sn}$ parameters in sample \#2 and \#4 have a larger standard deviation because the Sn composition is lower, such that less Sn-Sn 1NN pairs are available for statistical analyses in each nanocube. The Gaussian fitting curve of each histogram is also shown in Fig.\ref{fig:SnSn_SRO_Histogram}. A notable distinction is that the MBE samples have a larger $\alpha^{1NN}_{Sn-Sn}$, indicating a stronger presence of Sn-Sn 1NN pairs. In particular, almost 90\%  of the nanocubes in MBE sample \#1 (63 out of 72) and 66\% of those in sample \#2 (66 out of 100) showed Sn-Sn 1NN SRO parameters greater than 1. The average $\alpha^{1NN}_{Sn-Sn}$ is 1.14 for both sample \#1 and \#2, in contrast to $\sim$1.01 for samples \#3 and \#4. A Welch’s t-test reveals a highly statistically significant difference in the mean Sn-Sn 1NN SRO parameters ($\alpha^{1NN}_{Sn-Sn}$) between the MBE vs. CVD samples (p$<$0.0001). As discussed earlier, the measured difference of $\sim$0.13 in $\alpha^{1NN}_{Sn-Sn}$ should be taken as the \textit{lower limit} considering atomic position perturbations, i.e., the actual difference should be $\geq$0.13.

\subsection{Impact of Sn-Sn SRO on GeSn bandgap}\label{subsec5}

\subsubsection{Experimental results}

\begin{figure}[H]\centering
    \center{\includegraphics[width=1.0\textwidth]
    {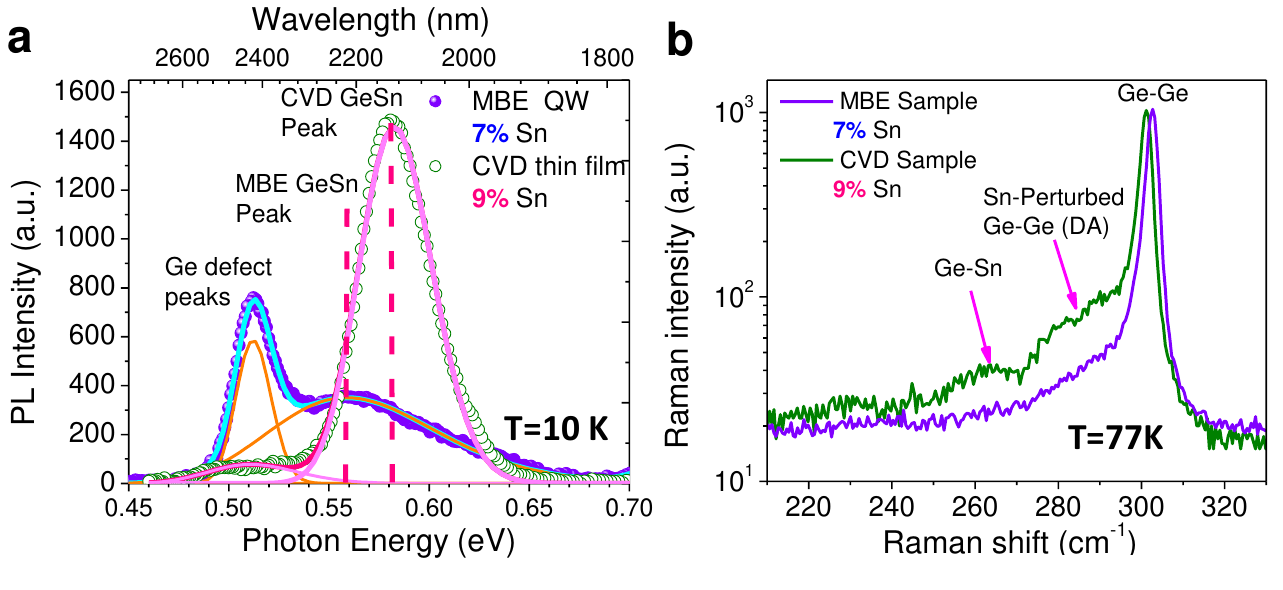}}
    \caption{ a) Comparison of PL spectra at 10 K between MBE QW Sample 2 (maximal 7 at.\% Sn) and a CVD thin film CVD GeSn thin film sample with 9 at.\% Sn (see Supporting Information for more details). The compressive strain in the GeSn layers is -1.02\% for the MBE sample and -1.18\% for the CVD sample, i.e. very similar. b) The corresponding Raman spectra at 77 K. The shoulder around 280 $cm^{-1}$, commonly known as "DA" peak in literature, is induced by Sn-perturbed Ge-Ge mode. A higher DA peak intensity indicates a reduction in Sn-Sn 1NN pairs \cite{corley2023local}.}
    \label{fig:PL}
\end{figure}

As theoretically predicted in \cite{cao2020short}, depletion of Sn-Sn 1NN leads to an increase in GeSn bandgap due to alleviation of local lattice dilation induced by large Sn atoms. Conversely, enrichment of Sn-Sn 1NN pairs would decrease the bandgap. The difference in Sn-Sn SRO parameters between MBE and CVD samples revealed in the previous section allows us to directly test this prediction for the first time. Fig. \ref{fig:PL}a compares the photoluminescence (PL) spectra measured at 10 K using 1064 nm laser excitation for Sample 2 (MBE QW, maximal 7 at.\% Sn \cite{Eldose_2025}) vs. a CVD GeSn thin film sample with 9 at.\% Sn (more details of this sample are available in Section 3 of the Supporting Information). We chose these two samples because they are fully strained on Ge buffer layers as revealed by reciprocal space mapping (see \cite{Eldose_2025} and Figure S3, respectively), thereby mitigating the impact of dislocation defects in GeSn. Low temperature PL avoids complications such as thermal broadening, which helps to better compare the peak positions. It also facilitates direct comparison with DFT band structure modeling at T=0 K.  Here the narrower peaks at 0.512 eV  originate from dislocations in the Ge substrate or the Ge buffer layer \cite{https://doi.org/10.1002/pssc.200675485,Eldose_2025}. The identical Ge dislocation peak positions from both samples indicate that the spectra are well calibrated. The broader peaks at 0.55-0.6 eV correspond to the band-to-band transitions in GeSn. The PL peak width of the MBE sample is broader due to the Sn composition variation shown in Fig. \ref{fig:APT_Tips}d. In fact, its PL peak width of $\sim$80 meV is largely consistent with $\sim$30 meV change in bandgap per 1 at.\% Sn composition difference reported in previous literature \cite{10.1063/1.4897272}. 

Reciprocal space mapping reveals similar compressive strain for these two samples, i.e. -1.02\% (MBE) vs. -1.18\% (CVD) (see \cite{Eldose_2025} and Figure S3, respectively). Furthermore, the comparison of Raman spectra at 77 K in Fig. \ref{fig:PL}b shows that the Ge-Ge peak of the MBE sample is indeed blueshifted by 1.8 $cm^{-1}$ compared to the CVD sample, consistent with 2 at.\%Sn compositional difference under similar strain (i.e. 0.95 $cm^{-1}$ blueshift per 1 at.\% decrease in Sn composition) \cite{10.1063/1.4943192}. Therefore, considering that the CVD sample has a higher Sn composition and no quantum confinement, one would expect a smaller bandgap from the 9 at.\% Sn CVD sample compared to the maximal 7 at.\% Sn MBE QW sample (Sample 2). 

Surprisingly, the PL spectra in Fig. \ref{fig:PL}a show exactly the \textit{opposite}: the PL peak of the 7 at.\% Sn MBE QW sample is redshifted by approximately 25 meV, or nearly 100 nm in wavelength, compared to the 9 at.\% CVD sample! This phenomenon clearly cannot be explained by composition (opposite trend), strain (almost the same), or quantum confinement in QWs (inducing blueshift instead of redshift). Thus, the only plausible explanation is that the significantly stronger presence of Sn-Sn 1NN in MBE samples overrides the composition and quantum confinement effects, resulting in a narrower bandgap despite a 2 at.\% lower Sn content. This interpretation is further supported by the aforementioned theoretical prediction in \cite{cao2020short} and the Raman spectra in Fig. \ref{fig:PL}b. The latter show that the Sn-perturbed Ge-Ge mode intensity (i.e. the "DA" peak) is much weaker in the MBE sample, indicative of stronger presence of Sn-Sn 1NN \cite{corley2023local} than its CVD counterpart in agreement with the APT analyses. Based on previous studies \cite{10.1063/1.4897272}, decreasing the Sn composition from 9 to 7 at.\% would \textit{widen} the bandgap by $\sim$60 meV. Since we actually observed a 25 meV \textit{narrower} bandgap in the 7 at.\% Sn MBE sample than that of the 9 at.\% Sn CVD sample, the larger Sn-Sn 1NN SRO parameter in the MBE samples must have reduced the direct bandgap by at least 85 meV. To the best of our knowledge, this is the first experimental observation of a significant impact of Sn-Sn SRO on the bandgap of GeSn semiconductor alloys.

\subsubsection{First-principles band structure calculations}

\begin{figure}[H]\centering
    \center{\includegraphics[width=1\textwidth]
    {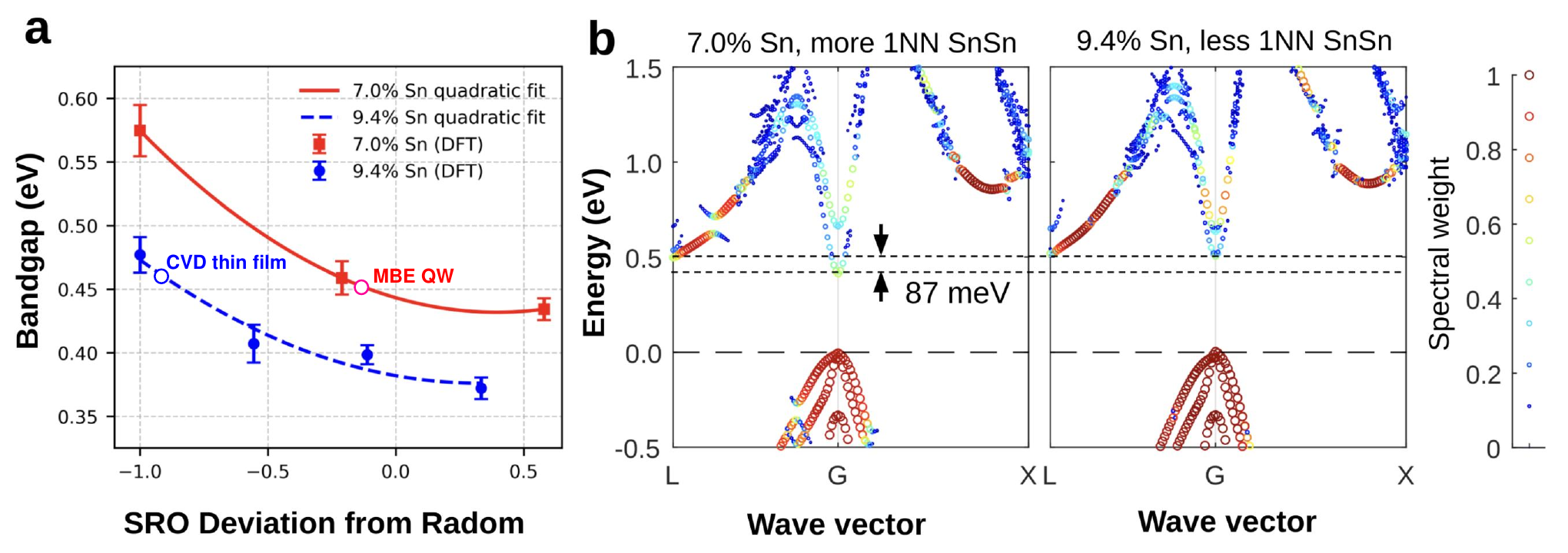}}
    \caption{Band structures. a) Bandgap values as a function of Sn-Sn 1NN SRO parameter deviation from random alloys, obtained from DFT calculations for GeSn alloys with 7.0 at.\% and 9.4 at.\% Sn at varying degrees of SRO. The bandgaps of the 7 at.\% Sn (MBE-grown) and 9 at.\% Sn (CVD-grown) samples, after correction for the strain effect, are also marked on the figure by the red and blue hollow circles, respectively. b) Unfolded band structures from DFT calculations for GeSn alloy of 7.0 at.\% with more preferred 1NN Sn-Sn (corresponding to SRO for the rightmost red square in (a)) and GeSn alloy of 9.4 at.\% Sn with less preferred 1NN Sn-Sn (corresponding to SRO for the leftmost blue circle in (a)).}
    \label{fig:bandgap_SRO}
\end{figure}

To confirm that SRO is responsible for the experimentally observed PL redshift, we calculate the band structures of 
GeSn alloys of 7.0 at.\% Sn and 9.4 at.\% Sn with different Sn-Sn 1NN SRO parameters (see Atomistic modeling). As shown in Fig. \ref{fig:bandgap_SRO}\textbf{a}, the calculated bandgaps for both Sn compositions indeed exhibit strong dependence on Sn-Sn 1NN SRO, with the bandgap increasing by $\sim$120 meV as the Sn-Sn 1NN SRO parameter decreases from that of random alloy to complete Sn-Sn 1NN depletion. We note that this dramatic difference in bandgap resulting from SRO is equivalent to 4 at.\% variation in Sn composition based on the previous literature \cite{10.1063/1.4897272}, highlighting the significant tunability of GeSn alloy bandgap through SRO engineering. More importantly, our calculation corroborates the experimental results, showing that a GeSn alloy of 7.0 at.\% Sn with a strong presence of Sn-Sn 1NN pairs can indeed exhibit a smaller bandgap than a 9.4 at.\% Sn GeSn alloy depleted of Sn-Sn 1NN pairs by as much as 87 meV (see Fig. \ref{fig:bandgap_SRO}\textbf{b}). 

We further map the experimentally measured bandgaps of the 7 at.\% Sn MBE sample and its 9 at.\% Sn CVD counterpart onto the theoretically modeled bandgap vs. Sn-Sn SRO parameter plots, as shown by the hollow circles in Fig. \ref{fig:bandgap_SRO}. Since the modeling is based on relaxed GeSn while the experimental data are measured from GeSn thin films with $\sim$-1$\%$ compressive strain, we corrected the strain effect using the deformation potential reported in \cite{10.1063/1.4943192} for $\sim$8$\%$ Sn composition. This analysis shows that the Sn-Sn 1NN SRO parameter of the 7 at. \% Sn MBE sample is higher than that of the 9.4 at. \% Sn CVD sample by $\sim$0.7. This result agrees with the Sn-Sn 1NN SRO parameter difference of $\geq$0.13 derived from APT data analyses, considering that APT tends to underestimate the difference in SRO parameters due to atomic position perturbations. Actually, referring to Fig. S2 in the Supporting Information, this underestimated difference in SRO parameters by $\sim$0.6 can be well explained by a Gaussian perturbation of atomic positions with a standard deviation of 3-4 \AA (i.e., the corresponding $\alpha^{1NN}_{Sn-Sn}$=0.82 in Fig. S2, compared to 0.22 without perturbation). Overall, comparisons between the experimental and theoretical analyses above indicate that increasing Sn-Sn 1NN SRO parameter by $\sim$0.7 decreases the bandgap by at least 85 meV, equivalent to the impact of increasing Sn composition by $\sim$3 at.\%. This significant SRO-induced bandgap modification represents $\sim$20\% of the material's entire bandgap ($\sim$500 meV for relaxed GeSn). 

\subsection{First-principles modeling to understand the SRO difference between MBE and CVD GeSn }\label{subsec5}

\begin{figure}[H]\centering
    \center{\includegraphics[width=0.6\textwidth]
    {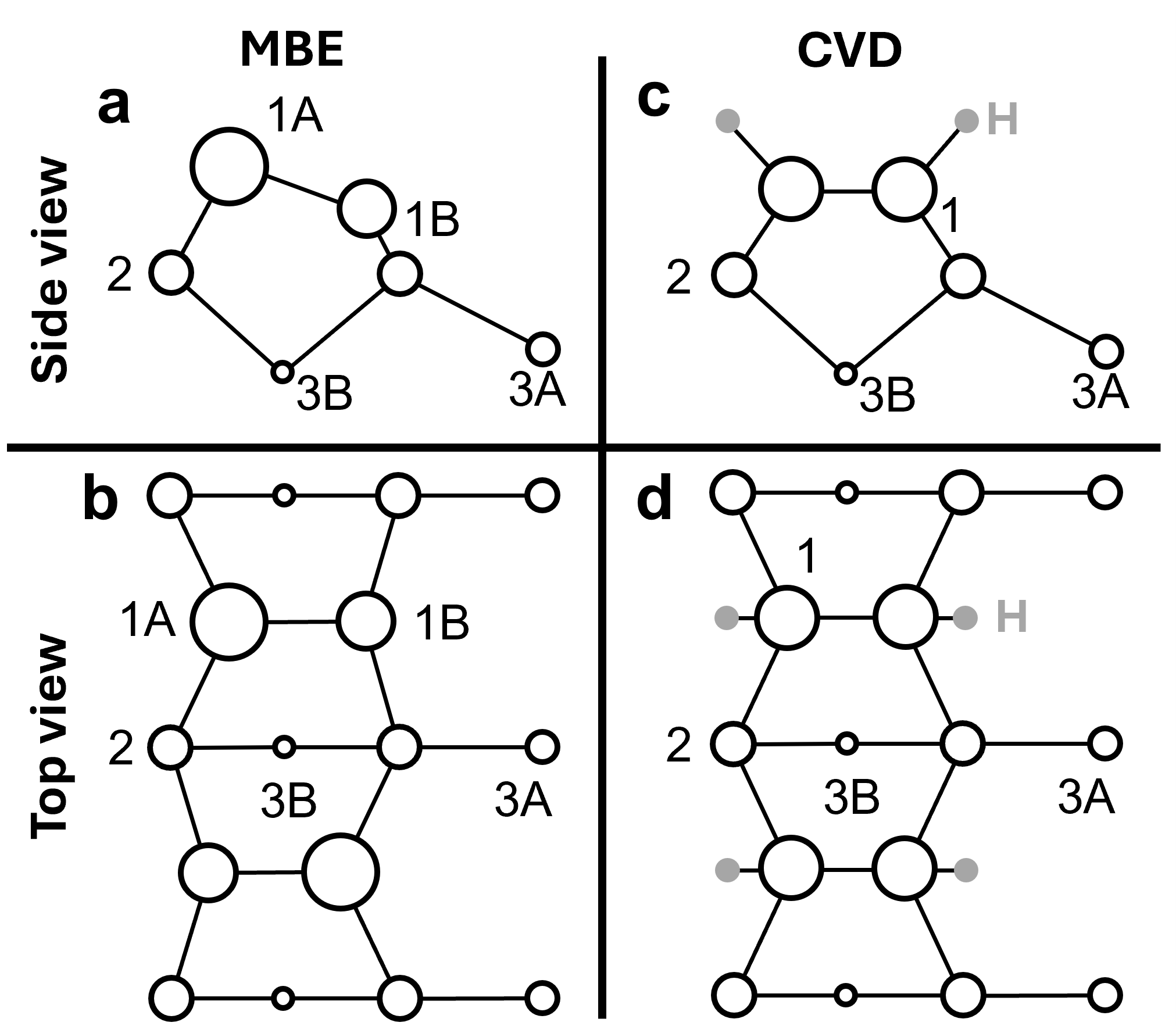}}
    \caption{The a) side and b) top view of atomic structures for the free surface with p(2$\times$2) reconstruction. The c) side and d) top view of atomic structures with mono-hydrogenated surface. The little grey circle represents the hydrogen atom and others represent Ge or Sn atoms. Different sizes of atoms represent different classes of equivalent sites. }
    \label{fig:DFT_Model}
\end{figure}

To understand the impact of growth method on SRO, we carry out DFT study on the energetics of Sn in the near-surface region during growth in the presence/absence of H (see the Atomistic modeling section), which sets a critical distinction between the CVD and MBE processes. Here we focus on a direct side-by-side comparison. In MBE growth without surfactant, the growth front is considered to be a free surface, as shown in Fig. \ref{fig:DFT_Model}a. Each atom on the surface holds two dangling electrons. One of these bonds is stabilized through the formation of a dimer with a neighboring surface atom. The two remaining dangling bonds from each dimer pair distributed to one side leads to the buckling of the dimers. To reduce the strain energy, the adjacent surface dimers are buckled in opposite way, leading to the p(2$\times$2) surface reconstruction \cite{zandvliet2003ge} shown in Fig. \ref{fig:DFT_Model}b. According to the symmetry, the atomic sites near the surface can be classified into 5 types as labeled in Fig. \ref{fig:DFT_Model}a. 

For chemical vapor deposition (CVD), the precursor and the carrier gas include hydrogen atoms, yielding a hydrogenated top surface. As shown in Fig. \ref{fig:DFT_Model}c, d, the hydrogen atoms passivate all the unpaired electrons in the surface dimer. As a result, the buckling of  dimers disappears and the 1A and 1B atomic sites in Fig. \ref{fig:DFT_Model}a become equivalent and relabeled as 1 in Fig. \ref{fig:DFT_Model}c.

\begin{figure}[H]\centering
    \center{\includegraphics[width=0.6\textwidth]
    {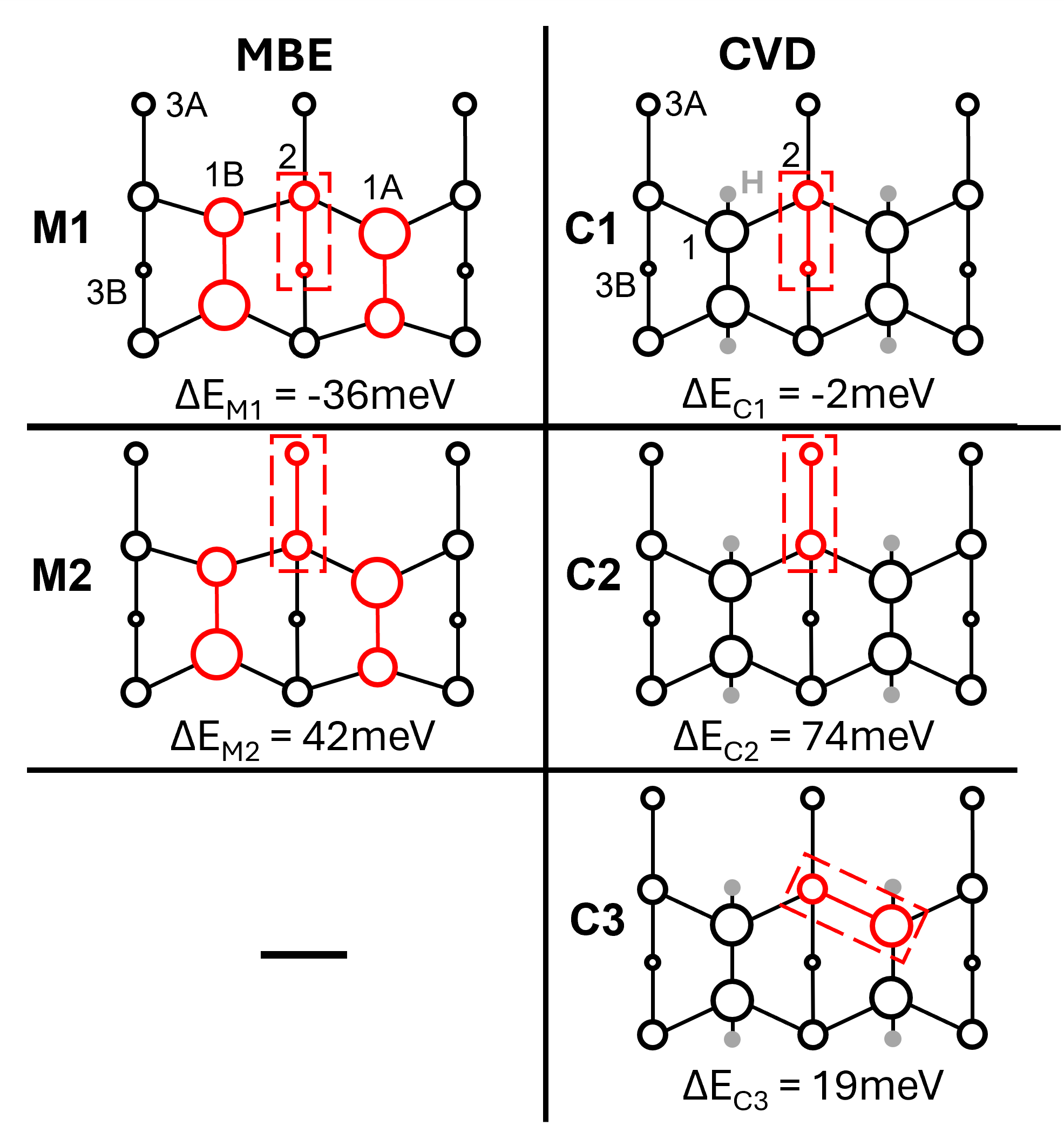}}
    \caption{The interaction between Sn atoms in 1$^{st}$ neighbor shell with different configurations for MBE (left) and CVD (right) near the top surface. The atomic site occupied by Sn is highlighted by red color and all other sites are Ge. The Sn-Sn dimer where $\Delta$E is calculated is highlighted by red dash box.  }
    \label{fig:DFT_FormationEnergy}
\end{figure}

During the growth process, Sn diffusion can be mainly assumed to occur between the surface and the sub-surface (see Atomistic modeling). Therefore, the interaction between Sn atoms in the top three layers can play an important role in the formation of SRO. Such an interaction between Sn atoms can be quantitatively characterized by the energy change ($\Delta$E) when two isolated Sn atoms are brought together to form a 1NN bond. The interaction is repulsive when $\Delta$E is positive and attractive when $\Delta$E is negative. %Since the energy of the system is different when a Sn atom occupies different class of atomic sites labeled in Fig. \ref{fig:DFT_Model}, the Sn atoms must occupy the same class of atomic sites for both isolated and 1NN cases to make the energy comparable. 
For MBE growth, Sn has a strong tendency to segregate to the top surface,\cite{johll2015influence} and previous experiment has shown that the growth front tends to have higher Sn concentration. \cite{wegscheider1992fabrication} Therefore we assume that the surface (1A and 1B sites) is fully occupied by Sn atoms. As a result, only the interaction between Sn in the 2nd and 3rd layers from the top surface should be considered. As shown in the left panel of Fig. \ref{fig:DFT_FormationEnergy}, all symmetry inequivalent sites for Sn-Sn dimers are labeled as M1 (between sites 2 and 3B) and M2 (between sites 2 and 3A). Energy calculations show that Sn atoms in the M1 configuration are attractive (negative $\Delta$E), whereas those in the M2 configuration are repulsive (positive $\Delta$E).

In contrast, for CVD growth where the surface is hydrogenated, Sn segregation can be suppressed.\cite{johll2015influence} Therefore, all possible configurations of Sn-Sn dimers within the top three layers should be considered, including configurations C1 and C2 which correspond to M1 and M2 in MBE growth, respectively. By comparing $\Delta E_{C1}$ to $\Delta E_{M1}$ or $\Delta E_{C2}$ to $\Delta E_{M2}$, we find a tendency that H atoms can enhance the repulsion between Sn atoms by $\sim$35 meV. Additionally, there is a new C3 configuration for Sn-Sn dimer between sites 1 and 2, which also shows repulsive interactions between Sn atoms. Therefore, the hydrogenated surface in CVD growth can lead to a less preference for Sn-Sn 1NN than MBE, driven by stronger repulsive interaction between Sn atoms. This finding from first-principles modeling supports the experimental results discussed earlier, showing a stronger presence of Sn-Sn 1NN pairs in MBE samples than CVD ones. 

\section{Discussion}\label{sec3}

Statistical analyses of APT data reveal that MBE GeSn samples show a stronger presence of Sn-Sn 1NN pairs than their CVD counterparts for both thin films and MQW cases, irrespective of the growth tools at different institutions. This comparison indicates that the $\alpha^{1NN}_{Sn-Sn}$ SRO parameters in MBE samples are higher than CVD ones by $\geq$0.13. The APT analysis is further supported by Raman spectroscopy, showing notably weaker DA peaks in MBE sample indicative of higher likelihood of Sn-Sn 1NN pairs. Corroborating these experimental results, first-principles modeling reveals that the reconstructed GeSn surface in vacuum during MBE growth tends to favor Sn-Sn 1NN at the growth front, as indicated by the negative $\Delta E_{M1}$ in Fig. \ref{fig:DFT_FormationEnergy}. In contrast, H passivation during CVD growth helps to reduce Sn-Sn 1NN pairs at the growth front. 

We also note that higher growth temperature tends to diminish Sn-Sn 1NN pairs at the surface growth front \cite{Liang2025}, since the magnitude of Sn-Sn 1NN interaction energy is on the order of tens of meV (up to $\sim$80 meV), which is comparable to the thermal energy $k_B T$ at elevated CVD growth temperatures (e.g., $\approx$54 meV at 350\degree C). Therefore, in addition to the difference in surface termination, the much higher CVD growth temperature (250-350 \degree C) also helps reduce the Sn-Sn 1NN SRO parameters compared to MBE growths (120-150 \degree C). 

Most importantly, we further experimentally demonstrate the significant impact of Sn-Sn 1NN SRO parameters on the bandgap of GeSn for the first time, confirming previous theoretical predictions \cite{cao2020short,chen2024intricate} that enhancement of Sn-Sn 1NN would shrink the bandgap of GeSn. Comparison of experimental results in Fig. \ref{fig:PL} with first-principles bandgap modeling in \ref{fig:bandgap_SRO} reveals that increasing Sn-Sn 1NN SRO parameter by $\sim$0.7 remarkably reduces the bandgap by $\geq$85 meV compared to random alloys, equivalent to the impact of substantially increasing Sn composition by $\sim$3 at.\%. These results establish SRO as a new and independent degree of freedom for semiconductor alloy band engineering beyond composition, strain, and quantum confinement.

Overall, the synergy between our theoretical and experimental analyses indicate that surface termination and growth temperature are two effective levers to control Sn-Sn 1NN SRO in GeSn alloys for band engineering. Similarly, it has been reported that surfactant can affect SRO in other systems such as GaAsSb and GaAsN.\cite{jiang2004effect,occena2018surfactant}. Furthermore, in the case of ternary SiGeSn alloys, SRO has been predicted to play an even more critical role in determining the band structures, potentially changing the bandgap by as much as 3x at the same chemical composition.\cite{jin2022coexistence,jin2023role} Therefore, we envision that extending surface termination and growth temperature control to SiGeSn will harness SRO for even more dramatic band engineering. Because this new SRO paradigm allows for band structure tuning without changing the composition or lattice constant, it opens the door to novel electronic and photonic device structures inaccessible to conventional approaches. An attractive example will be addressing the long-standing challenge of lattice-matched type-I QWs on Si for high-performance photonic devices by leveraging SRO-induced electronic heterostructures and QWs \cite{Jin2024}.

Last but not least, we note that these insights on SRO control during GeSn synthesis apply directly to post-growth SRO modifications. In fact, very lately the impact of temperature on SRO has been utilized to modify the bandgap of GeSn nanowires via post-growth annealing \cite{GeSnNWSRO}. Conventional phase change materials (PCMs) for photonics and electronics experience a density and volume change of 5\% to 10\% upon phase transitions,  which critically limits cycling durability in PCM switches and non-volatile memories \cite{PCMlimitation, PCMNVM}. In contrast, SRO transitions can shift the bandgap by $\sim$20\% at zero volume change because they only alter the atomic neighborhood without changing the macroscopic chemistry, crystal structure, or lattice constant. Engineering SRO transitions can thus enable major breakthroughs for reconfigurable electronics and photonics by overcoming a fundamental durability bottleneck. 

\section{Conclusion}\label{sec7}

In summary, we analyze Sn-Sn SRO in thin-film GeSn alloys grown by MBE vs. CVD and find a stronger presence of Sn-Sn 1NN in MBE-grown samples than in CVD-grown ones. This finding aligns with theoretical modeling, which shows that surface hydrogen termination and elevated growth temperature in CVD reduce the preference for Sn-Sn 1NN pairs. The comparison also reveals that surface termination and growth temperature are the primary levers for SRO control during GeSn synthesis. Furthermore, synergistic experimental and theoretical analyses reveal that increasing the Sn-Sn 1NN SRO parameter by $\sim$0.7 decreases the bandgap by at least 85 meV. This significant redshift represents $\sim$20\% of the material's direct bandgap ($\sim$500 meV), matching the impact of a substantial 3 at.\% increase in total Sn composition. These results establish SRO as a new degree of freedom for band engineering beyond composition, strain, and quantum confinement. Because this new paradigm only alters the atomic neighborhood without changing macroscopic chemistry or lattice constant, it opens the door to novel electronic and photonic device structures inaccessible to conventional approaches. These insights also apply directly to post-growth SRO transitions, offering a strategy to overcome the volume change bottleneck of existing PCMs for highly durable reconfigurable electronics and photonics. Therefore, we envision that harnessing SRO will unlock new device mechanisms for the post-Moore's Law era.

\section{Experimental Section}
\textit{APT measurement conditions:} APT data were collected using pulse laser energies ranging from 5 to 75 pJ. Detailed measurement conditions for the four samples are provided in Table S1 in the supporting information. The reconstructions were performed using Cameca IVAS software.

\textit{Statistical analysis:} As a physics-informed statistical method, this Poisson-KNN approach incorporates the number of KNNs for a given atom in the corresponding crystal structure, as well as the detection efficiency of the experimental APT data. The number of A-B pairs in each true KNN shell is also weighted summed by the A-B pairs in the nominal KNN shells. Then $P^{KNN}_{A-B}$ is calculated as the ratio between the number of A-B pairs and the total number of A-all pairs in the statistically reconstructed KNN shell. To reach statistical significance, nanocubes of 5 nm $\times$ 5 nm $\times$ 5 nm  were chosen in this paper to calculate local SRO parameters in each nanocube and map the spatial distribution of SRO. Approximately 2000 atoms in each nanocube are sufficient for statistical analysis. The number of nanocubes analyzed for the four samples were 72, 100, 250, and 200, respectively, providing sufficient data for statistical analysis.

\textit{Programming:} The Poisson-KNN method programs were written using Python.\cite{liu20223d} APT data was imported by Python package ``apt-tools''.\cite{Branson2016Git}

\section{Atomisitic modeling}
\textit{Band structure calculation:} Structure models for GeSn alloys of 7.0 at.\% Sn and 9.4 at. \% Sn with different degrees of SRO are obtained using Monte Carlo sampling at various temperatures (300 K, 600K, and 1500 K) utilizing a highly accurate and efficient machine-learning potential for GeSn alloys \cite{chen2024intricate} based on the neuroevolution potential architecture for multi-component alloy systems \cite{song2024general}. For band structure calculations, all structures are further optimized using the local density approximation (LDA). A \(2\times2\times2\) Monkhorst-Pack \(k\)-points grid \cite{Monkhorst:1976cv} with a plane-wave cutoff energy of 300 eV is employed. The conjugate-gradient algorithm is applied for structural relaxation during each energy calculation, with convergence criteria set at \(10^{-4}\) eV and \(10^{-3}\) eV for electronic and ionic relaxations, respectively. To improve the bandgap accuracy, we use the modified Becke-Johnson (mBJ) exchange potential \cite{Tran:2009kk} in combination with LDA correlation, as implemented in the Vienna Ab initio Simulation Package (VASP) \cite{kresse1996efficient}, with the $c$-mBJ parameter set to 1.2. This approach accurately predicts bandgaps for Si, Ge, and $\alpha$-Sn, showing good agreement with experimental data while being computationally more efficient than hybrid functionals or GW methods \cite{Tran:2009kk,Eckhardt:2014gz,Polak:2017fh,cao2020short,jin2023role,chen2024intricate}. We employ the spectral weight approach \cite{Popescu:2010jd,Rubel:2014fv} to unfold the band structures back into the first Brillouin Zone of the diamond cubic structure, using the code \textit{fold2bloch} \cite{Rubel:2014fv}. Relativistic effects, specifically spin-orbit coupling, are included in the band structure calculation, as they are crucial for accurately reproducing the band structures of Ge and $\alpha$-Sn \cite{Eckhardt:2014gz,Polak:2017fh}. 

%\textcolor{red}{Here we should still add a brief summary of band structure calculations}. All using DFT calculations (see method details in Ref. \cite{chen2024intricate}).

\textit{Growth modeling:} First-principles calculations were conducted on the energy of Sn-Sn in the top three layers at the growth front with and without H termination, with Perdew-Burke-Ernzerhof (PBE) functional in the Vienna Ab initio Simulation Package (VASP) \cite{kresse1996efficient} based on the projector augmented wave method.\cite{blochl1994projector} A 16.4 \AA \ $\times$ 16.4 \AA \ $\times$ 34.7 \AA \ supercell with 13 atomic layer of Ge along (001) direction, which is the same as the growth directions of the samples, and 17 \AA \ vacuum along z axis is applied. The x-y lattice constant is fixed to the fully relaxed lattice constant of Ge bulk ($\sim$5.78 \AA). The top layer is applied to investigate the interaction between Sn atoms and the lower surface is fully passivated by adding two H atoms to each Ge atom. The H and lowest Ge layer is fixed to avoid additional strain effect. An energy cutoff of 300 eV and a 4 $\times$ 4 $\times$ 1 Monkhorst-Pack k-points grid are chosen for structural relaxation, combined with the convergence criteria of $10^{-6}$ and $5\times10^{-3}$ eV for electronic and ionic relaxations, respectively. During the growth process, with a low defect density, Sn atoms in the Ge bulk exhibit a high diffusion barrier ($\sim$3.2 eV), even when mediated by vacancies.\cite{chroneos2008vacancy} Thus, their positions are hard to change under the standard growth temperature ($\sim$150 \degree C for MBE and $\sim$300 \degree C for CVD). However, previous experiments have shown that the diffusion barrier for Sn near the top surface of GeSn and SiGeSn alloy can be less than 1 eV,\cite{li2014characteristics,taoka2015non} which allows the Sn atoms to rearrange themselves within this region. Assuming that the Sn diffusion mainly happens between the surface and the sub-surface, the interaction between Sn atoms in the top three layers can play an important role in the formation of SRO. In this calculation, we focus on the interaction between Sn atoms within the 1NN shell.

%%%%%%%%%%%%%%%% REFERENCES %%%%%%%%%%%%%%%

\clearpage % Clear all remaining figures and tables then start a new page

% The list of references goes after the main text and before the acknowledgements
% When preparing an initial submission, we recommend you use BibTeX, like this:
%
\bibliography{reference} % for a file named science_template.bib
\bibliographystyle{sciencemag}

% After the paper has completed peer review and been revised ready for acceptance,
% you should comment out the lines above and copy-paste the contents of your .bbl
% file here instead. This will help ensure that our conversion software works correctly.
% Remember to re-run BibTeX first - check the timestamp!
%
% Example of the first three entries copy-pasted from science_template.bbl:
%
%\begin{thebibliography}{1}
%
%\bibitem{example}
%A.~N. {Author}, An example reference. \emph{Journal of Improbable Research}
%  \textbf{1}, 67 (2020).
%
%\bibitem{example2}
%F.~M. {Surname}, S.~{Author}, A second example. \emph{Interesting Research
%  Letters} \textbf{32}, 897 (2019).
%
%\bibitem{example_preprint}
%P.~{One}, P.~{Two}, P.~{Three}, {An unpublished preprint}. \emph{preprint}
%  (2021), arXiv:2101.12345.
%
%\end{thebibliography}

%%%%%%%%%%%%%%%% ACKNOWLEDGEMENTS %%%%%%%%%%%%%%%

\section*{Acknowledgments}
The Poisson-KNN method of analyzing SRO from the APT data had been developed under the support of the Air Force Office of Scientific Research under the award number FA9550-19-1-0341. The APT tip preparation and data collection of the GeSn samples provided by University of Arkansas and University of Delaware, the DFT computational modeling of MBE vs. CVD growths, and the supercell modeling of APT data have been supported by $\mu$-ATOMS, an Energy Frontier Research Center funded by the U.S. Department of Energy (DOE), Office of Science, Basic Energy Sciences (BES), under the award DE-SC0023412.

\paragraph*{Author contributions:}
J.L. conceived this work. S.L. calculated SRO from APT data. Y.L. and S.Z. performed DFT total-energy calculations. S.C., X.J. and T.L. generated theoretical GeSn supercell data and computed the impact of SRO on the bandstructure of GeSn. A.C. and X.W. analyzed data. J.B., I.B., C.C., A.A. conducted APT data acquisition and reconstruction. H.Z., N.E., O.C., Y.Z., D.B., G.S. and S.Y. provided GeSn samples.

\paragraph*{Competing interests:}
There are no competing interests to declare.

\paragraph*{Data and materials availability:}
The data and code that support the findings of this study are available from the corresponding author upon reasonable request.

%%%%%%%%%%%%%%%% SUPPLEMENT LIST %%%%%%%%%%%%%%%

% List the contents of your Supplementary Materials, including the numbers of any
% supplementary figures, tables, external data files etc. and any references that are
% cited only in the supplement. In this example, refs. 7-8 are cited only in the supplement.
% Fill out your numbers accordingly and delete any lines that aren't applicable.
\subsection*{List of supplementary materials}

\paragraph*{Table S1:} APT measurement conditions.
CVD sample for photoluminescence (PL) studies.

\paragraph*{Fig. S1:}  $\alpha^{1NN}_{Sn-Sn}$ versus (a) Laser energy. (b) Ge+/Ge++ ratio. (c) Sn+/Sn++ ratio. Sample \#1-2 (red) were grown by MBE and sample \#3-4 (blue) were grown by CVD.

\paragraph*{Fig. S2:}  (a) Poisson-KNN derived Sn-Sn 1NN SRO parameter $\alpha^{1NN}_{Sn-Sn}$ vs. the standard deviation of Gaussian atomic position perturbation of theoretically simulated $Ge_{0.75}$$Sn_{0.25}$ SRO and random alloy supercells using machine-learning potential. Red curve: a random $Ge_{0.75}$$Sn_{0.25}$ alloy with true $\alpha^{1NN}_{Sn-Sn}$ = 1. Purple curve: a $Ge_{0.75}$$Sn_{0.25}$ alloy with SRO, $\alpha^{1NN}_{Sn-Sn}$ = 0.22. Ten different configurations are generated by randomly removing 50\% of atoms for each case to match the detection efficiency of APT measurements, followed by random perturbation of atomic positions based on Gaussian distribution at a given standard deviation. The SRO parameter is then averaged over these 10 randomly perturbed supercells. The error bars are smaller than the size of the symbols. (b) Normalized histogram of reconstructed true 1NN interatomic distance in the 4 samples studied in this paper. The standard deviation is 1.04, 0.98, 1.03, 1.06 \AA \ for sample \#1-4, respectively.

\paragraph*{Fig. S3:} (a) XRD and (b) RSM data of the 9.0 at.\% CVD sample for PL studies
%%%%%%%%%%%%%%%% END OF MAIN TEXT %%%%%%%%%%%%%%%

\end{document}